\documentclass[conference]{IEEEtran}
\IEEEoverridecommandlockouts
\usepackage{cite}
\usepackage{amsmath,amssymb,amsfonts}
\usepackage{algorithm}
\usepackage{algpseudocode}
\usepackage{graphicx}
\usepackage{textcomp}
\usepackage{xcolor}
\usepackage{lipsum}

\def\BibTeX{{\rm B\kern-.05em{\sc i\kern-.025em b}\kern-.08em
    T\kern-.1667em\lower.7ex\hbox{E}\kern-.125emX}}
\begin{document}

\title{Fast and Interactive Byzantine Fault-tolerant Web Services via Session-Based Consensus Decoupling}


\author{
\IEEEauthorblockN{
    Ahmad Zaki Akmal, Azkario Rizky Pratama, and Guntur Dharma Putra\IEEEauthorrefmark{1} \\
    Universitas Gadjah Mada, Indonesia \\
    ahmad.zaki.akmal@mail.ugm.ac.id, \{azkario, gdputra\}@ugm.ac.id
    }
\thanks{\IEEEauthorrefmark{1}Corresponding author.}
}

\maketitle

\begin{abstract}
Byzantine fault-tolerant (BFT) web services provide critical integrity guarantees for distributed applications but face significant latency challenges that hinder interactive user experiences. We propose a novel two-layer architecture that addresses this fundamental tension between security and responsiveness in BFT systems. Our approach introduces a session-aware transaction buffer layer (Layer 2) that delivers immediate feedback to users through consensus simulation, while periodically committing batched operations to a fully Byzantine fault-tolerant consensus layer (Layer 1). By separating interactive operations from consensus finalization, our system achieves responsive user experiences of under 200ms, while maintaining strong BFT security guarantees. We demonstrate the efficacy of our architecture through a supply chain management implementation, where operators require both immediate feedback during multi-step workflows and tamper-proof record keeping. Our evaluation shows that our Layer 2 operations perform four times faster than the Layer 1 counterpart, while substantially preserving the end-to-end transaction integrity. Our approach enables BFT applications in domains previously considered impractical due to latency constraints, such as metaverse environments, where users require both responsive interaction and guaranteed state consistency.
\end{abstract}


\section{Introduction}
\label{sec:introduction}

The concept of metaverse represents a shift in digital interaction, evolving beyond isolated virtual experiences toward persistent, interconnected 3D spaces where users can socialize, work, and engage in commerce~\cite{encyclopedia2010031, rony_e-commerce_2024}. The metaverse vision aligns naturally with Web3 principles, where blockchain technology enables decentralized ownership, transferable digital assets, and trustless transactions~\cite{hatami_survey_2024}. As these interconnected economies mature, enabling users to trade assets with real-world values~\cite{yang_web30_2023}, the underlying web services must evolve beyond traditional centralized architectures. Open metaverses particularly benefit from Byzantine fault tolerance due to their decentralized nature~\cite{rajawat_enhancing_2023}, ensuring consensus and transaction validity without relying on trusted intermediaries, a necessity when handling valuable digital assets in environments without central authority.

Byzantine Fault-tolerant (BFT) systems operate correctly even when up to one-third of the nodes act maliciously~\cite{lamport_byzantine_1982, castro_practical_2002}, making them crucial for metaverse applications where digital assets represent real economic value. However, current BFT web services face significant latency challenges that hinder interactive user experiences. The state-of-the-art DeWS (Decentralized Web Services) introduces approximately one second of latency for a 15-node system configuration~\cite{ramachandran_dews_2023}. Such delays create substantial friction in interactive environments where users expect near-instantaneous feedback, representing a fundamental tension between security guarantees and interactive responsiveness.

To illustrate this challenge, consider a supply chain workflow with multiple verification steps involving several stakeholders: suppliers, warehouse operators, and couriers. Each package undergoes verification, quality checks, and labeling before delivery. Warehouse operators require instant feedback during each processing step, while business management needs to ensure correctness and establish trust between stakeholders. The challenge becomes evident during high-volume periods when operators cannot afford delays waiting for consensus verification between steps, yet the business cannot compromise on creating an auditable and tamper-proof record of all activities. Similar requirements exist in healthcare, financial services, e-government, and metaverse environments, where participants need both high responsiveness during operations and guaranteed integrity of the final record. In these multi-step workflows, even small delays at each verification stage compound exponentially, potentially bringing operations to a standstill while participants wait for consensus confirmation. This operational bottleneck not only impacts productivity but also undermines user adoption of otherwise secure distributed systems.

Current approaches to this challenge typically force a binary choice between security and performance. Traditional centralized systems achieve low latency but sacrifice fault tolerance and require trusted authorities, making them unsuitable for cross-organizational workflows or decentralized environments. Conversely, existing BFT systems prioritize security but impose consensus overhead on every operation, resulting in cumulative delays that render them impractical for interactive applications. This fundamental limitation has restricted BFT adoption to scenarios where occasional high latency is acceptable, leaving a significant gap for applications requiring both security guarantees and responsive user interactions.

In this paper, we propose a two-layered web service architecture to address the latency challenge while maintaining Byzantine fault tolerance. The first layer (L1) conducts full BFT consensus across a distributed network of nodes, while the second layer (L2) implements a novel session-based transaction buffer that provides immediate feedback to users through consensus simulation. We introduce the concept of \textit{sessions}, where related operations are grouped and processed through the responsive L2 simulation before being committed as a batch to the more secure but slower L1 consensus. This approach significantly reduces both perceived latency and the frequency of slower L1 consensus operations.

Our work offers these key contributions:
\begin{itemize}
    \item We introduce \textit{a two-layer architecture for BFT web services} that separates interactive simulation from consensus finalization, enabling responsive user experiences while maintaining security guarantees.
    
    \item We develop \textit{a transaction buffering mechanism} that groups related operations into sessions, simulates their execution with up-to-date state data, and commits them as atomic units, reducing perceived latency while preserving transactional integrity.
    
    \item We provide \textit{a fully functional proof-of-concept} for our proposed solution through a supply chain scenario that demonstrates its practical feasibility along with a detailed performance evaluation.
\end{itemize}

The remainder of this paper is organized as follows. Section~\ref{sec:related-work} reviews related work on BFT systems and Layer 2 technologies. Section~\ref{sec:system-model} presents our system architecture, while Section~\ref{sec:implementation} describes our implementation details. We conclude our work in Section~\ref{sec:conclusion}.

\section{Related Work}
\label{sec:related-work}

Byzantine Fault Tolerance (BFT) addresses the challenge of maintaining system correctness despite arbitrary failures or malicious behavior in some components. In web services, BFT is particularly important when operations span multiple organizations with different trust boundaries~\cite{lamport_byzantine_1982}.

Practical Byzantine Fault Tolerance (PBFT)~\cite{castro_practical_2002} made BFT implementations feasible by reducing communication complexity from exponential to polynomial. Zyzzyva~\cite{kotla_zyzzyva_2008} further improves performance by lowering latency, involving clients to assist in validations. However, PBFT, Zyzzyva, and their derivatives still face scalability challenges, particularly in high-throughput interactive applications.

Several approaches have applied BFT concepts to web services, including Thema~\cite{merideth_thema_2005}, BFT-WS~\cite{zhao_bft-ws_2007}, WebBFT~\cite{berger_webbft_2018}, and the most recent, DeWS~\cite{ramachandran_dews_2023}. Thema and BFT-WS implement Byzantine fault tolerance for Web Services while maintaining compatibility with standard SOAP and WSDL protocols, but they rely on centralized middleware architectures, which can become potential bottlenecks or single points of failure. WebBFT later improved on these ideas by making BFT services accessible to browser-based clients. It supported real-time collaborative features like publish-subscribe updates and worked with modern web technologies such as WebSockets~\cite{berger_webbft_2018}. Despite these advances, WebBFT faced several challenges. It introduced noticeable performance overhead from browser-side cryptography and JSON handling, and was vulnerable to denial-of-service attacks caused by malicious clients triggering repeated leader changes. These issues limited its scalability and robustness in adversarial environments.

A key similarity among these earlier efforts is that they achieve BFT by reaching consensus on the computation result of a single web server or through the inclusion of centralized components, thus lacking full computation replication across nodes. In contrast, Ramachandran et al. argue that replicating the complete computation at each node is necessary to ensure safety in the presence of Byzantine faults, a principle that underpins their design of DeWS~\cite{ramachandran_dews_2023}.

DeWS implements a brand new interaction model: \textit{request-compute-consensus-log-response}. All operations undergo replication and consensus validation before responses are returned to clients, ensuring that responses are agreed upon by a quorum of nodes. This approach provides strong integrity and auditability guarantees but introduces significant latency—approximately 935ms with 15 nodes even in a Docker container network, which represents an ideal environment compared to real-world deployments~\cite{ramachandran_dews_2023}. This latency creates a fundamental tension between Byzantine fault tolerance and interactivity, making DeWS challenging to apply in scenarios requiring frequent user interactions or real-time feedback.

Layer 2 blockchain solutions address similar scalability challenges by offloading transaction processing from the main chain. Techniques such as rollups and sidechains enable higher throughput and lower latency while maintaining security by periodically anchoring state to the layer 1 chain~\cite{mandal_investigating_2023, thibault_blockchain_2022}.

Session management, the grouping of related operations into logical units, has been explored in distributed systems primarily for user authentication and state tracking~\cite{calzavara_measuring_2021}. However, its potential for optimizing consensus operations remains largely unexplored in BFT web services.

\begin{figure}[t]
    \centering
    \includegraphics[width=0.85\linewidth]{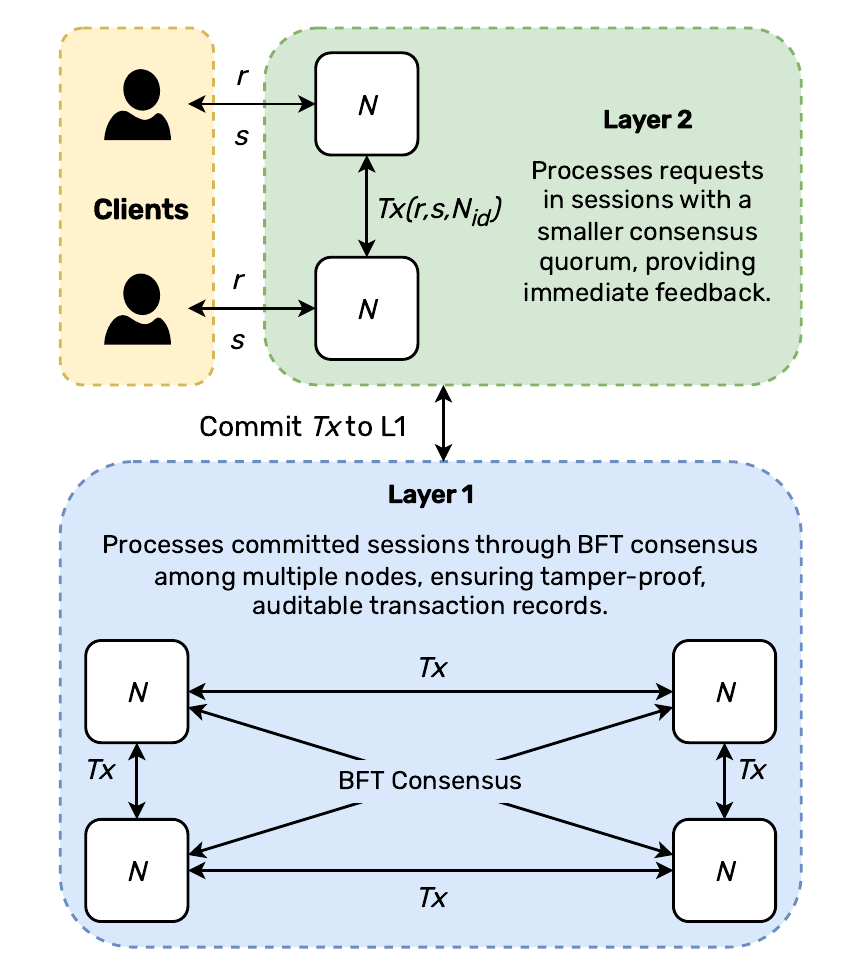}
    \caption{Our two-layer architecture for BFT web services, showing client interactions with Layer 2 nodes and eventual session commitment to the Layer 1 BFT consensus network.
    }
    \label{fig:system_model}
\end{figure}

While these previous works have advanced Byzantine fault tolerance in web services, they collectively fail to address the fundamental trade-off between interactive responsiveness and strong BFT guarantees. The absence of a solution that maintains both low-latency interactivity and strong BFT guarantees represents a significant gap in the literature. Our two-layer architecture directly addresses this gap by separating the concerns of immediate feedback and consensus validation, enabling responsive user experiences without sacrificing the integrity guarantees of Byzantine fault tolerance.

\section{System Model and Architecture}
\label{sec:system-model}

In this section, we describe our proposed solution that addresses the latency challenges in distributed web services while maintaining Byzantine fault tolerance.

\subsection{System Architecture}
Figure~\ref{fig:system_model} illustrates the architecture of our proposed system. The system employs a dual-layer approach, each consisting of interconnected nodes with specific responsibilities. As shown in the diagram, every node in both layers incorporates these core components:

\begin{itemize}
\item Consensus Engine: Manages all consensus-related functions, including transaction verification, BFT consensus execution, block commitment, and ledger storage. This component also handles peer-to-peer (P2P) communication with other nodes within the same layer.
\item Web Server: Provides an HTTP interface exposing endpoints that enable clients to interact with the system through standardized APIs.
\item Service Registry: This component maps API routes to their corresponding handlers.
\item Database: Stores session data, transaction history, and application-specific information required for system operation.
\end{itemize}

These components work together to form a cohesive system that achieves both Byzantine fault tolerance and responsive performance.

\subsection{Layer 1 (L1): Byzantine Fault-Tolerant Foundation}

The first layer implements a design inspired by DeWS, where each node represents a domain operated by a distinct organization participating in the application ecosystem~\cite{ramachandran_dews_2023}. Unlike traditional BFT research that focuses on abstract state machine replication, our approach necessarily integrates business verification with consensus processes—a design choice required for cross-organizational applications where business rule enforcement cannot rely on trusted intermediaries.

L1 serves as the primary foundation for BFT by implementing full BFT consensus protocols. Nodes in L1 maintain identical service registries and execute consensus only when receiving commit requests from L2 nodes. Each validator independently verifies transaction validity by applying the same service handler as the transaction originator and comparing response equivalence.

To satisfy BFT requirements, L1 must contain at least four nodes, which is the minimum necessary to tolerate one Byzantine fault, following the standard rule that a BFT system requires at least $3f+1$ nodes to tolerate $f$ Byzantine faults. In addition to the standard node components, L1 nodes maintain an immutable blockchain ledger, providing tamper-proof storage of confirmed transactions.

\begin{algorithm}[t]
\caption{Unified Consensus Process (Both L1 and L2)}
\label{algorithm:validate-tx-process-proposal}
\begin{algorithmic}[1]

\Procedure{ValidateTransaction}{transaction}
    \State \textbf{Layer 1:} Verify transaction is well-formed and properly signed
    \State \textbf{Layer 2:} Check transaction against session state and business rules
    
    \State Decrypt and parse transaction content
    
    \If{transaction format is invalid}
        \State \Return FALSE \Comment{Malformed transaction}
    \EndIf
    
    \If{transaction violates state constraints}
        \State \Return FALSE \Comment{Transaction is invalid}
    \EndIf
    
    \State Execute transaction against current state
    
    \State \Return TRUE \Comment{Transaction is valid}
\EndProcedure

\Procedure{ProcessProposal}{proposedBlock}
    \State \textbf{Layer 1:} Validators independently execute transactions and compare results
    \State \textbf{Layer 2:} Simulates validation by re-executing operations against local state
    \If{results from execution differ from proposed results}
        \State \Return REJECT \Comment{Byzantine behavior detected}
    \Else
        \State \Return ACCEPT \Comment{Proposed transactions is valid}
\EndIf
\EndProcedure



\end{algorithmic}
\end{algorithm}

\subsection{Layer 2 (L2): Interactive Transaction Layer}

The second layer represents the core innovation of our architecture. L2 functions as a transaction buffer, performing simulation consensus and batching operations for efficient commitment to L1. Each node in L2 employs the same service registry and service handlers both in the web server and consensus components.
Unlike L1, this layer is not required to fulfill the strict BFT node count requirements; instead, it inherits fault tolerance properties from the underlying L1 network.

The separation between layers enables L2 to prioritize fast responsiveness over native BFT, effectively optimizing system performance for interactive applications while still validating transactions according to system rules. Furthermore, L2 is designed to operate with a minimal number of nodes, allowing consensus simulation to complete faster and reducing the latency introduced during transaction processing.

\subsection{System Flow}

The process begins when a client sends a request to initiate a new session with a Layer 2 node. The receiving node, becoming the transaction originator, processes the request by using its service registry to identify the appropriate service handler. The handler processes the request locally and stores the session in the node's database. After processing, the service handler generates a response $s$, which together with the original request $r$ and the node's identifier $N_{id}$ forms a transaction $Tx(r, s, N_{id})$.

\begin{figure}[t]
    \centering
    \includegraphics[width=1\linewidth]{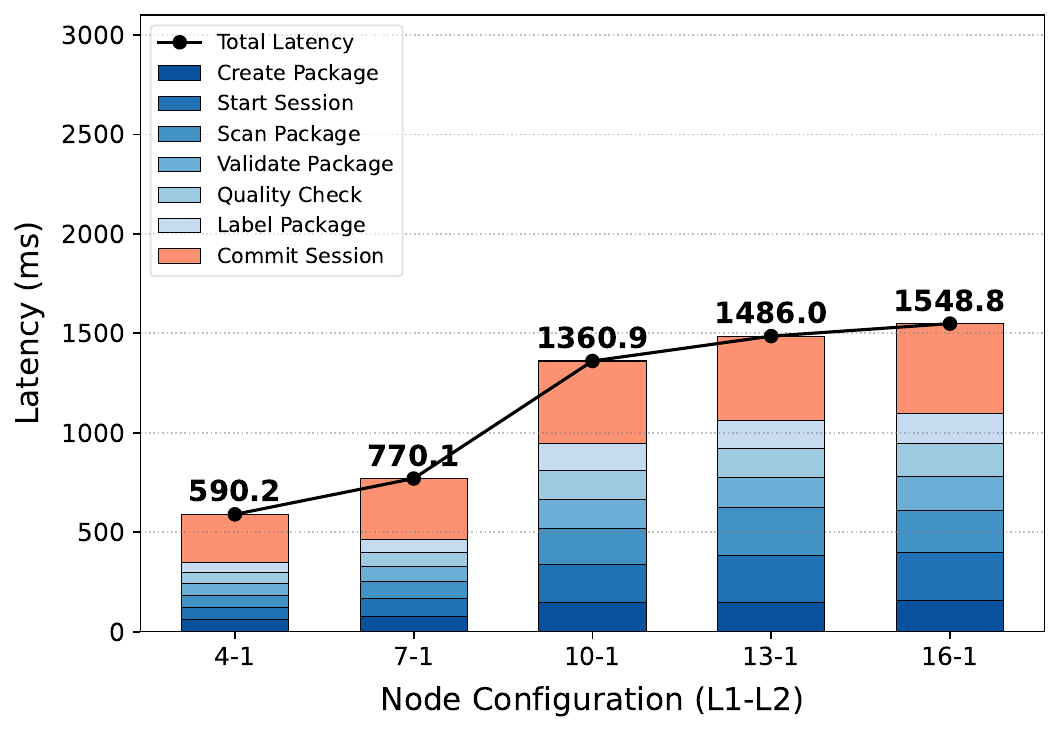}
    \caption{Detailed endpoint latency breakdown for configurations with a single L2 node, showing individual endpoint contributions to total workflow latency.}
    \label{fig:stacked-endpoints-l2-1}
\end{figure}

The transaction originator forwards this transaction to the L2 consensus layer, which performs simulation consensus by replicating the operation across other L2 nodes. As shown in Algorithm~\ref{algorithm:validate-tx-process-proposal}, each participating node processes the request through its own service registry and service handlers to verify response consistency. During the Process Proposal phase, nodes execute transactions independently and reject proposals if the results differ, detecting Byzantine behavior. If the deployment has only a single L2 node, it still executes consensus validation steps internally to ensure compliance with network rules.


After session creation, clients can execute a series of related operations within the same session context. Each subsequent request follows the same processing pattern as the initial session creation, with the important distinction that each operation is validated against the session's current state, following the validation logic in Algorithm~\ref{algorithm:validate-tx-process-proposal}.

The L1 layer operates a full BFT consensus mechanism. Similar to the L2 process, the transaction originator on L1 first replicates the complete session across validators. It generates a transaction containing all operations and submits it to the L1 consensus layer. The BFT consensus algorithm requires validation from at least $\lceil \frac{2n+1}{3} \rceil$ nodes to consider the transaction valid; otherwise, it is rejected.

Upon reaching consensus, the L1 network returns the result to the L2 node along with blockchain reference data such as block height and transaction hash. The L2 node then updates the session status accordingly, propagates this update to other L2 peers, and delivers the final result to the client. For verification purposes, clients can query transaction status either through an L2 node or directly from the L1 network, depending on the implementation.

This two-layer approach combines the immediate responsiveness of simulation consensus with the security guarantees of BFT consensus, creating a system that is both interactive and trustworthy.

\begin{figure}[t]
    \centering
    \includegraphics[width=1\linewidth]{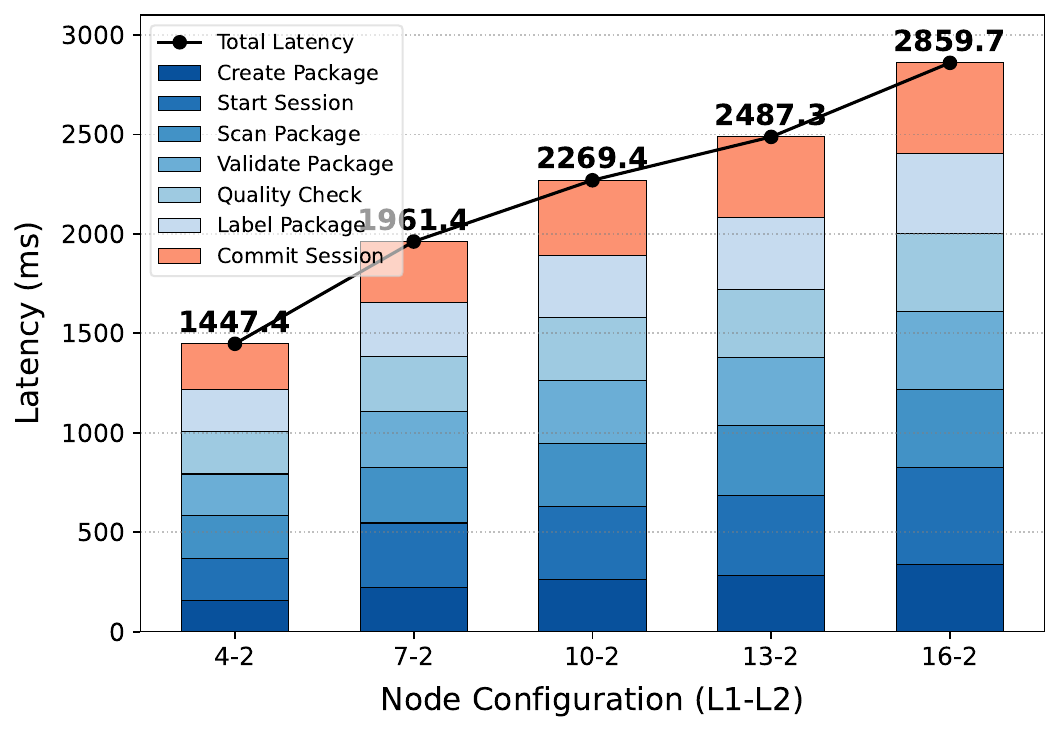}
    \caption{Detailed endpoint latency breakdown for configurations with dual L2 nodes, showing individual endpoint contributions to total workflow latency.}
    \label{fig:stacked-endpoints-l2-2}
\end{figure}

\section{Performance Evaluation}
\label{sec:implementation}

Our proof-of-concept implementation demonstrates a supply chain management system that represents an ideal scenario where both high responsiveness and strong security guarantees are the required attributes. Supply chain operations demand fast and interactive processing to maintain throughput while simultaneously requiring tamper resistance to prevent fraud across organizational boundaries. As such, our use case demonstrates the need for session fault tolerance, step-by-step feedback without premature commitment, and a tamper-proof audit trail. The implemented workflow consists of five sequential stages: 
(1) session initiation, where an operator begins processing a package; (2) package scanning, which identifies and retrieves expected contents; 
(3) validation of the package's digital signature to authenticate its origin; 
(4) quality control inspection; and 
(5) shipping label generation and courier assignment. Once completed, the entire session can be committed as an atomic unit to the Byzantine fault-tolerant L1, demonstrating how our dual-consensus architecture effectively balances responsive user experience with strong security guarantees in multi-step business processes. It should be noted that while this workflow represents a simplified version of real-world supply chain operations, it captures the essential characteristics and challenges that our system is designed to address.

\subsection{Proof-of-Concept Implementation}


We utilized CometBFT as the consensus engine, providing customizable and robust BFT using the Tendermint consensus algorithm~\cite{buchman_latest_2019, buchman_revisiting_2022} through its \textit{Application BlockChain Interface} (ABCI). We implemented both the web server and service registry components in Go programming language to ensure optimal interoperability with CometBFT's native codebase.

Our primary goal is to enhance interactivity and responsiveness through the L2 implementation, so our evaluation focuses primarily on latency metrics. We define latency as the time it takes from request initiation to response reception, calculated as $t_{\text{res}} - t_{\text{req}}$, where $t_{\text{res}}$ represents the response reception timestamp and $t_{\text{req}}$ represents the request initiation timestamp.

\begin{figure}[t]
    \centering
    \includegraphics[width=1\linewidth]{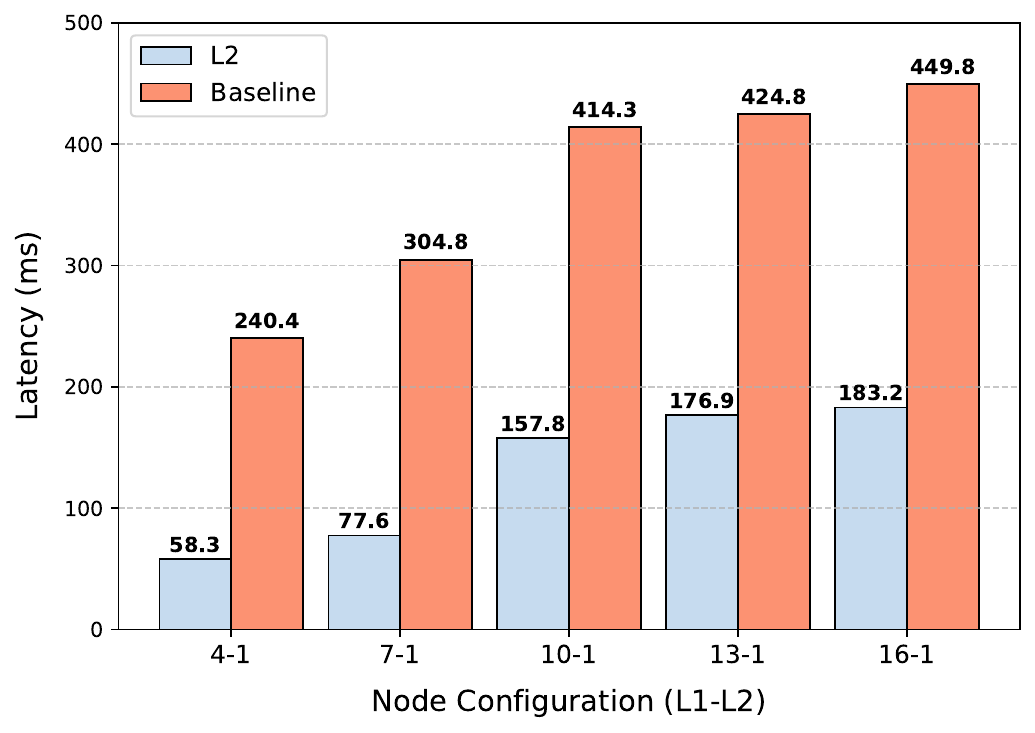}
    \caption{Response time comparison between L1 commitment operations and L2 simulation with L2=1 configurations, showing the significant decrease in latency at about 2.4-4.1× faster.}
    \label{fig:baseline-comparison-l2-1}
\end{figure}

We conducted comprehensive measurements for each step in the workflow sequence: from session creation through to blockchain commitment. To facilitate multiple test iterations, we implemented a package creation step that generates unique package identifiers for each test run. This approach enabled us to execute measurements systematically across varying node configurations. For each complete workflow, we performed 100 iterations to ensure statistical significance and reliability of our results.

For clarity in presenting our results, we use the notation $L1$-$L2$, where $L1$ represents the number of nodes in the BFT Layer 1 network and $L2$ represents the number of nodes in the simulation Layer 2 network. We systematically tested configurations with 1 and 2 L2 nodes while varying L1 nodes across five different cluster sizes: 4, 7, 10, 13, and 16. These specific node counts were selected to represent increasing Byzantine fault tolerance thresholds from 1 to 5, calculated according to the standard formula $n = 3f+1$~\cite{lamport_byzantine_1982}, where $n$ is the total number of nodes and $f$ is the number of tolerable Byzantine faults.


\subsection{Experimental Results}

Our evaluation results demonstrate the significant performance advantages of the L2 session batching approach across various BFT cluster configurations.

For configurations with a single L2 node, we observed clear performance patterns as the number of L1 nodes increased. L2 operations (Create Package, Start Session, Scan Package, Validate Package, Quality Check, and Label Package) maintained relatively low latencies even as the BFT cluster size grew. Figure~\ref{fig:stacked-endpoints-l2-1} provides a detailed breakdown of individual endpoint latencies across all tested configurations with a single L2 node. The complete workflow latency increased from 590.2ms with the 4-1 configuration to 1548.8ms with the 16-1 configuration. This increase is primarily attributable to the growing complexity of achieving consensus across larger BFT networks, where additional network communication and validation steps are required for each additional node.

\begin{figure}[t]
    \centering
    \includegraphics[width=1\linewidth]{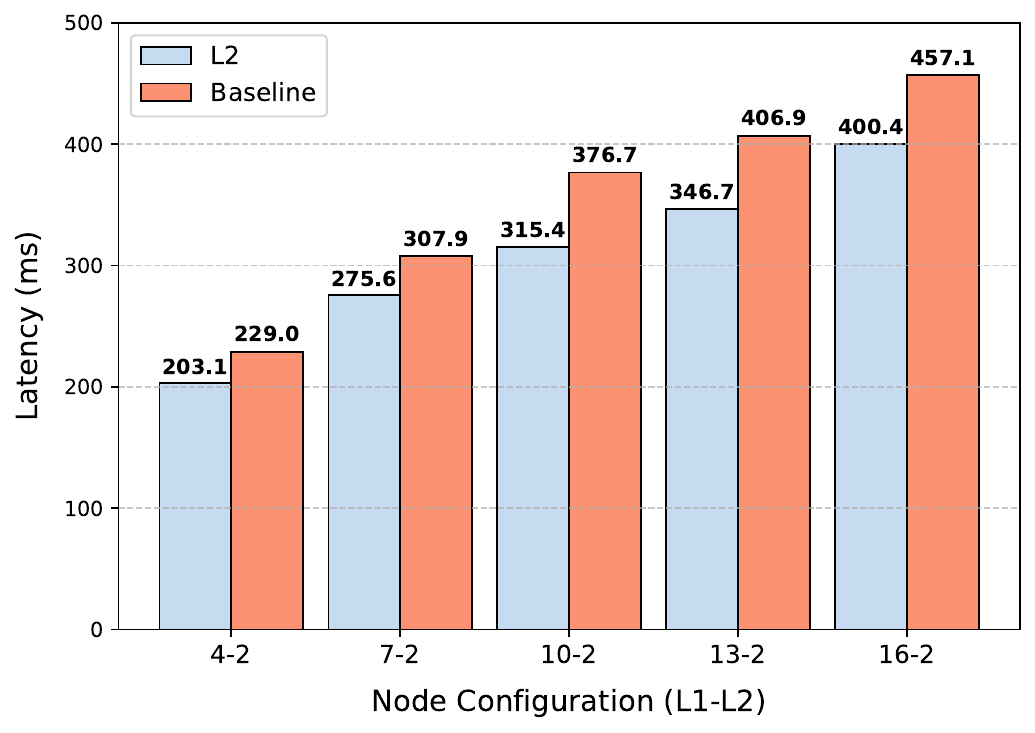}
    \caption{Response time comparison between L1 consensus commit operations and L2 simulation with L2=2 configurations, showing reduced performance advantages due to coordination overhead between dual L2 nodes.}
    \label{fig:baseline-comparison-l2-2}
\end{figure}

To better represent realistic consensus simulation, we tested configurations with two L2 nodes. Adding a second L2 node introduces additional coordination overhead but provides a more robust simulation environment. The endpoint-specific breakdown shows that complete workflow execution time ranges from 1447.4ms with the 4-2 configuration to 2859.7ms with the 16-2 configuration. This significant increase reflects the added complexity of coordinating two L2 nodes while simultaneously scaling the BFT L1 cluster, where inter-node communication overhead becomes more pronounced.

The direct performance comparison between L1 and L2 operations provides clear evidence of our approach's effectiveness. Looking at the comparison in Figure~\ref{fig:baseline-comparison-l2-1}, individual L2 operations averaged 58.3ms with the 4-1 configuration compared to 240.4ms for L1 consensus operations, while with the 16-1 configuration, L2 operations averaged 183.2ms compared to 449.8ms for L1 operations. Single L2 node configurations demonstrate a significant latency decrease, with L2 operations approximately 2.4 to 4.1 times faster than L1 commits across all tested cluster sizes. The performance improvement is most pronounced in smaller configurations (4.1× faster for 4-1) and remains substantial even with larger clusters (2.4× faster for 16-1).

In contrast, Figure~\ref{fig:baseline-comparison-l2-2} reveals that dual L2 node configurations show markedly reduced performance advantages due to coordination overhead between the two L2 nodes. With the 4-2 configuration, L2 operations averaged 203.1ms compared to 229.0ms for L1 operations, while scaling to the 16-2 configuration increased L2 operation latency to 400.4ms compared to 457.1ms for L1 operations. The performance improvement drops to only 1.1 to 1.2 times faster than L1 commits, with L2 latency approaching L1 latency across all configurations. This demonstrates that the coordination overhead between dual L2 nodes nearly eliminates the speed benefits of the L2 layer, making single L2 configurations significantly more effective for performance-oriented applications.

The experimental results validate our two-layer approach: Layer 2 operations consistently achieve sub-second latencies suitable for interactive use, while the more expensive BFT consensus operations are consolidated into a single commit phase.
The substantial improvement in responsiveness enables interactive workflows that would otherwise be impractical with direct L1 consensus for each operation.
While the latencies we observed are higher than those reported for traditional \textit{request-compute-response} operations in DeWS (approximately 16-18ms for basic POST/GET operations)~\cite{ramachandran_dews_2023}, the increased responsiveness in complex multi-step workflows justifies this trade-off. Our approach provides significant benefits in terms of session management, state consistency, and interactive user experience across the two-layer architecture.

Our findings suggest that smaller L1 configurations with a single L2 node offer the best balance of performance and fault tolerance for most applications. However, for scenarios requiring higher Byzantine fault tolerance thresholds, configurations with 10 to 16 L1 nodes remain viable, with total workflow latencies still acceptable for supply chain and similar applications. The dual L2 node configurations, while theoretically providing more accurate simulation, introduce significantly higher latencies without proportional improvements in L2 operation responsiveness. This suggests that for applications prioritizing performance, either a single L2 node or direct L1 operations would be preferable to a multi-node L2 setup, as the added complexity does not yield substantial benefits in our tested scenarios.

\section{Conclusion}
\label{sec:conclusion}
This paper presented a two-layer architecture with a session-aware transaction buffer that addressed the challenge of providing interactive experiences in Byzantine fault-tolerant (BFT) web services. Our approach decoupled latency-sensitive operations from BFT consensus processes, maintaining sub-200ms response times for interactive operations even with 16-node BFT clusters. By buffering operations and deferring consensus until session completion, we achieved both the responsiveness needed for modern applications and the security guarantees of Byzantine fault tolerance. Our approach opened possibilities for BFT systems in domains previously impractical due to latency constraints, particularly for multi-step workflows like supply chain management.

\section*{Acknowledgment}
\noindent This work was supported by Hibah Penelitian Fundamental (PFR), Ministry of Higher Education, Science, and Technology Indonesia, grant number 048/E5/PG.02.00.PL/2024 - 2679/UN1/DITLIT/PT.01.03/2024.

\bibliography{references}{}
\bibliographystyle{IEEEtran}


\end{document}